**High Frequency Haplotypes are Expected Events, not Historical Figures**

Elsa G. Guillot[1] and Murray P. Cox[1,*]

[1] Statistics and Bioinformatics Group, Institute of Fundamental Sciences, Massey University, Palmerston North, New Zealand

[*] Corresponding author

Contact Details:

Institute of Fundamental Sciences

Massey University

Private Bag 11 222

Palmerston North 4442

New Zealand

Phone: +64-6-356 9099 x84747

Fax: +64-6-350 5682

Murray P. Cox: m.p.cox@massey.ac.nz

Elsa G. Guillot: elza.guillot@gmail.com

**Abstract**

Cultural transmission of reproductive success states that successful men have more children and pass this raised fecundity to their offspring. Balaresque and colleagues found high frequency haplotypes in a Central Asian Y chromosome dataset, which they attribute to cultural transmission of reproductive success by prominent historical men, including Genghis Khan. Using coalescent simulation, we show that these high frequency haplotypes are consistent with a neutral model, where they commonly appear simply by chance. Hence, explanations invoking cultural transmission of reproductive success are statistically unnecessary.

Cultural transmission of reproductive success states that successful men have more children and pass this raised fecundity on to their offspring. Observed in modern human populations from genealogies and surname studies[1], in a genetic setting cultural transmission of reproductive success would cause particular male lines to dominate on the Y chromosome. Balaresque and colleagues[2] examined a Y chromosome dataset from Central Asia to determine whether they could reconstruct historic instances of this behavior. Screening 8 microsatellites on the Y chromosome in 5,321 Central Asian men (distribution in Figure 1), they identified 15 haplotypes that are carried by more than 20 men (grey bars). The authors described these haplotypes as 'unusually frequent,' but did not provide any statistical support for this statement. These lineages were subsequently connected by the authors to prominent historical figures, including Genghis Khan and Giocangga.

However, in any given haplotype frequency distribution, a number of haplotypes are expected to occur at high frequency simply by chance. In neutrally evolving systems, haplotype frequency distributions follow a Zipfian power law[3]: most lineages are carried by only a few men (Figure 1, left side), while a small number of lineages are carried by many men (Figure 1, right side). The Y chromosome distribution observed by Balaresque and colleagues closely follows such a power law, thus providing strong preliminary evidence that their Y chromosome dataset may be selectively neutral.

To more explicitly test whether the observed high frequency haplotypes are actually unusually frequent, we simulated genetic data under the standard coalescent, a neutral model that does not include cultural transmission of reproductive success. We modeled the evolution of 5,321 Y chromosomes, each carrying 8 fully linked microsatellites, to match the observed data. The code for these simulations, including full details of parameter values, is available online (http://elzaguillot.github.io/Allele-Frequency-Spectrum-simulations) [4].

Simulations were first run across a sweep of $\theta$ values to find the best match with the power law distribution observed in the Central Asian Y chromosome dataset. The least squares fit between observed and simulated distributions was minimized at $\theta = 131$. In one million simulations run at this value, we found that 27.2% of the simulations contained at least 15 haplotypes carried by more than 20 men, thus illustrating that high frequency haplotypes like those observed among Central Asian Y chromosomes are relatively common, even when cultural transmission of reproductive success is not acting. The Y chromosome haplotype frequency distribution observed by Balaresque and colleagues falls within the 95% confidence intervals of our simulations (Figure 1, red shading).

The most parsimonious explanation is therefore that the high frequency haplotypes observed by Balaresque and colleagues in Central Asia are simply expected chance events. While we strongly encourage further research into cultural transmission of reproductive success, no statistical evidence has been presented to show that this process has acted on this particular dataset of Central Asian Y chromosomes. As no additional evidence is presented in support of proposed links to famous historical men, these haplotypes instead most likely reflect the chance proliferation of random male lines, probably from historically unrecorded, culturally undistinguished, but biologically lucky Central Asian men.

## Software Availability

Latest source code for allele frequency spectrum simulations:

http://elzaguillot.github.io/Allele-Frequency-Spectrum-simulations

Archived source code as at the time of publication:

http://doi.org/10.5281/zenodo.29888

License:

Lesser GNU Public License 3.0 https://www.gnu.org/licenses/lgpl.html

## Author Contributions

EGG conceived the study and carried out the research. EGG and MPC designed the experiments and wrote the manuscript. Both authors were involved in the revision of the draft manuscript and have agreed to the final content.

## Competing Interests

No competing interests were disclosed.

## Grant Information

The authors declared that no grants were involved in supporting this work.

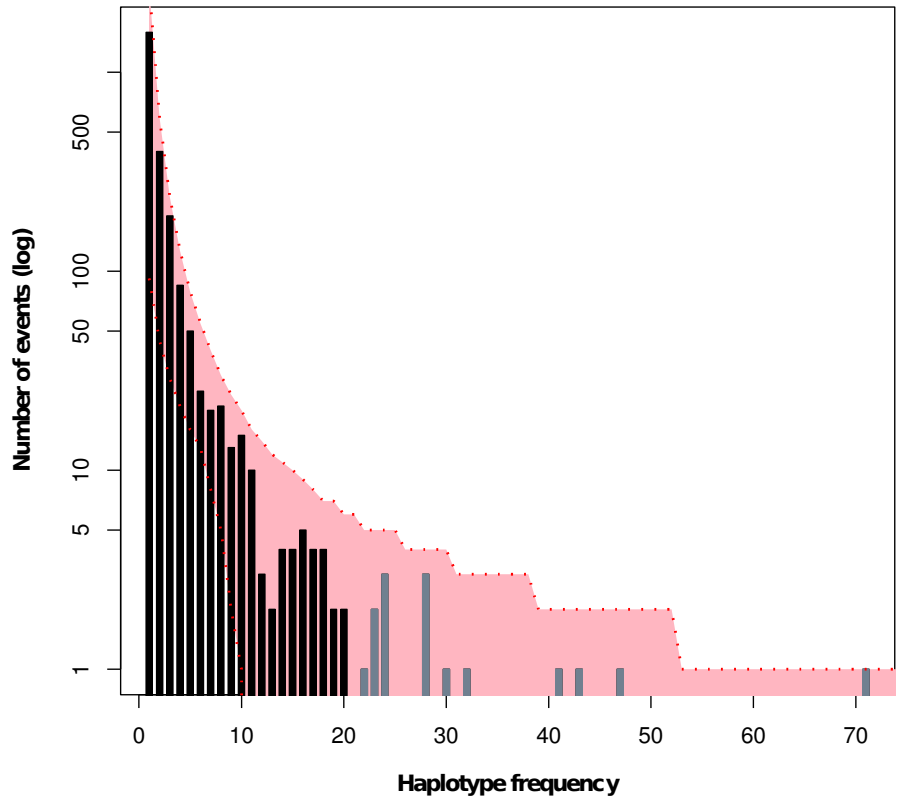

**Figure 1 Microsatellite haplotype frequency distribution.** The distribution (black and grey bars) is identical to Figure 2 of Balaresque *et al* [2]. Grey bars indicate the 15 haplotypes that Balaresque and colleagues describe as 'unusually frequent.' Red shading indicates the 95% confidence intervals of haplotype frequencies from one million simulations under a fitted neutral model. All of the high frequency haplotypes (grey bars) fall within these 95% confidence bounds.